\documentclass[prl,twocolumn,superscriptaddress]{revtex4-1}
\usepackage{amssymb}
\usepackage{amsfonts}
\usepackage{bbm}
\usepackage{mathrsfs}
\usepackage{graphics,graphicx,epsfig,bm,amsmath,amsthm,amssymb}
\usepackage{bm}
\usepackage{bbm}
\usepackage{longtable}
\usepackage{multirow}
\usepackage{array}
\usepackage{color}
\usepackage{cases}
\usepackage{booktabs }
\usepackage[usenames,dvipsnames]{xcolor}

\usepackage{xcolor}
\usepackage[none]{hyphenat}
\usepackage{float}
\usepackage{amsmath}
\usepackage[a4paper,colorlinks=true,
linkcolor=blue,citecolor=blue,
pdfauthor={ },
pdftitle={ },
pdfsubject={ },
pdfkeywords={ }]{hyperref}

\begin{document}
\title{Experimental Constraints on Exotic Spin-Dependent Interactions by a Magnetometer with Ensembles of Nitrogen-Vacancy Centers in Diamond}

\author{Hang Liang}
\affiliation{CAS Key Laboratory of Microscale Magnetic Resonance and School of Physical Sciences, University of Science and Technology of China, Hefei 230026, China}
\affiliation{CAS Center for Excellence in Quantum Information and Quantum Physics, University of Science and Technology of China, Hefei 230026, China}

\author{Man Jiao}
\affiliation{CAS Key Laboratory of Microscale Magnetic Resonance and School of Physical Sciences, University of Science and Technology of China, Hefei 230026, China}
\affiliation{CAS Center for Excellence in Quantum Information and Quantum Physics, University of Science and Technology of China, Hefei 230026, China}

\author{Yue Huang}
\affiliation{CAS Key Laboratory of Microscale Magnetic Resonance and School of Physical Sciences, University of Science and Technology of China, Hefei 230026, China}
\affiliation{CAS Center for Excellence in Quantum Information and Quantum Physics, University of Science and Technology of China, Hefei 230026, China}

\author{Pei Yu}
\affiliation{CAS Key Laboratory of Microscale Magnetic Resonance and School of Physical Sciences, University of Science and Technology of China, Hefei 230026, China}
\affiliation{CAS Center for Excellence in Quantum Information and Quantum Physics, University of Science and Technology of China, Hefei 230026, China}

\author{Xiangyu Ye}
\affiliation{CAS Key Laboratory of Microscale Magnetic Resonance and School of Physical Sciences, University of Science and Technology of China, Hefei 230026, China}
\affiliation{CAS Center for Excellence in Quantum Information and Quantum Physics, University of Science and Technology of China, Hefei 230026, China}

\author{Ya Wang}
\affiliation{CAS Key Laboratory of Microscale Magnetic Resonance and School of Physical Sciences, University of Science and Technology of China, Hefei 230026, China}
\affiliation{CAS Center for Excellence in Quantum Information and Quantum Physics, University of Science and Technology of China, Hefei 230026, China}

\author{Yijin Xie}
\affiliation{CAS Key Laboratory of Microscale Magnetic Resonance and School of Physical Sciences, University of Science and Technology of China, Hefei 230026, China}
\affiliation{CAS Center for Excellence in Quantum Information and Quantum Physics, University of Science and Technology of China, Hefei 230026, China}

\author{Yi-Fu Cai}
\affiliation{CAS Key Laboratory for Research in Galaxies and Cosmology, Department of Astronomy, University of Science and Technology of China, Hefei 230026, China}
\affiliation{School of Astronomy and Space Science, University of Science and Technology of China, Hefei 230026, China}

\author{Xing Rong}
\email{xrong@ustc.edu.cn}
\affiliation{CAS Key Laboratory of Microscale Magnetic Resonance and School of Physical Sciences, University of Science and Technology of China, Hefei 230026, China}
\affiliation{CAS Center for Excellence in Quantum Information and Quantum Physics, University of Science and Technology of China, Hefei 230026, China}

\author{Jiangfeng Du}
\email{djf@ustc.edu.cn}
\affiliation{CAS Key Laboratory of Microscale Magnetic Resonance and School of Physical Sciences, University of Science and Technology of China, Hefei 230026, China}
\affiliation{CAS Center for Excellence in Quantum Information and Quantum Physics, University of Science and Technology of China, Hefei 230026, China}

\date{\today}

\begin{abstract}
Improved constraints on exotic spin-dependent interactions are established at the micrometer scale by a magnetometer with ensembles of nitrogen-vacancy (NV) centers in diamond. A thin layer of NV electronic spin ensembles is utilized as the sensor, and a lead sphere is taken as the source of the nucleons. The exotic spin-dependent interactions are explored by detecting the possible effective magnetic fields by the sensor. Stringent bounds on an exotic parity-odd spin- and velocity-dependent interaction are set within the force range from 5 to 500 $\mu$m. The upper limit of the corresponding coupling constant, $g_A^eg_V^N$, is improved by more than three orders of magnitude at 330 $\mu$m. Improved constraints of $P, T$-violating scalar-pseudoscalar nucleon-electron interactions, are established within the force range from 6 to 45 $\mu$m. The limit of the corresponding coupling constant, $g_S^Ng_P^e$, is improved by more than one order of magnitude at 30 $\mu$m. Our result shows that a magnetometer  with a NV ensemble can be a powerful platform for probing exotic spin-dependent interactions.

\end{abstract}

\maketitle

New light bosons as dark matter candidates have attracted attention in astronomy and particle physics recently \cite{Dailey2021,Vermeulen2021,Afach2021}.
Among the most well-motivated new light bosons, axion addresses the strong CP mysteries \cite{peccei1977cp,weinberg1978new}, hierarchy problem \cite{Graham2015}, and is a generic prediction of string theories \cite{Svrcek2006, Gorghetto2021}.
The exotic spin-dependent interactions mediated by new bosons were first proposed by Moody and Wilczek \cite{Moody1984} and have been analyzed according to polarization and relative velocity between fermions \cite{dobrescu2006spin, Fadeev2019}.  The exotic spin-dependent interactions characterized by scalar, pseudoscalar, vector and axial-vector couplings lead to effective magnetic fields applied to polarized spins, which enables us to search for exotic spin-dependent interactions via precision measurement experiments in laboratory \cite{safronova2018search}.

Experimental searches for exotic spin-dependent interactions have been conducted using various kinds of sensors, such as the electron-spin polarized torsion pendulum \cite{Heckel2008}, superconducting quantum interference device (SQUID) \cite{Ni1999}, and spin-exchange-relaxation-free (SERF) atomic magnetometer \cite{Kim2019}. Recently, single nitrogen-vacancy (NV) centers in diamond have been developed as single-spin magnetometers \cite{Maze2008} to search for exotic spin-dependent interactions at micrometer scales \cite{Rong2018, Jiao2021PRL, Rong2018PRL}. An ensemble-NV-diamond magnetometer has been suggested to be utilized to conduct the search for exotic spin-dependent interactions \cite{Jiao2021PRL} due to its superior magnetic field sensitivity \cite{Taylor2008, Xie2021,Zhang2021,Barry2017,Barry2020}. Here we experimentally conducted searches for exotic spin-dependent interactions by an ensemble-NV-diamond magnetometer.

In this work, we focus on two types of exotic interactions between polarized electron spin and unpolarized nucleon, which are described as \cite{dobrescu2006spin},

\begin{equation}
V_{AV}(\textbf{r}) =  g_A^eg_V^N\frac{\hbar}{4 \pi}( \frac{e^{{-\frac{r}{\lambda}}}}{r})\bm{\sigma} \cdot \bm{v} ,
\label{VexoticVA}
\end{equation}

\begin{equation}
V_{SP}(\textbf{r}) =  g_S^N g_P^e\frac{\hbar^2}{8 \pi m_e} (\frac{1}{\lambda r}+\frac{1}{r^2}) e^{{-\frac{r}{\lambda}}} \bm{\sigma} \cdot \bm{e_r} ,
\label{VexoticSP}
\end{equation}
where $g_A^e$ ($g_V^N$) is the axial-vector (vector) coupling constant of new bosons to electrons (nucleons),  and $g_P^e$ ($g_S^N$) is the pseudoscalar  (scalar) coupling constant. $\lambda =\hbar/mc $ is the force range with $m$ being the mass of the hypothetical particle and $c$  the speed of light,  $\hbar$ is the reduced Planck's constant. $m_e$ is mass of the electron, $\bm{\sigma}$ is Pauli vector of the electron spin,  $\textbf{r}$ is the displacement vector between the electron and nucleon,  $r=|\textbf{r}|$ and $ \bm{e_r}=\textbf{r}/r$, $\bm{v}$ is the relative velocity. These interactions induce effective magnetic fields sensed by the electron spins of the NV centers,

\begin{equation}
\textbf{B}_{{eff,~AV}}(\textbf{r})=  \frac{g_A^eg_V^N}{2 \pi \gamma_e} ( \frac{e^{{-\frac{r}{\lambda}}}}{r}) \bm{v} ,
\label{BVA}
\end{equation}
\begin{equation}
\textbf{B}_{{eff,~SP}}(\textbf{r})= g_S^N g_P^e\frac{\hbar}{4 \pi m_e\gamma_e} (\frac{1}{\lambda r}+\frac{1}{r^2}) e^{{-\frac{r}{\lambda}}}   \bm{e_r} ,
\label{BSP}
\end{equation}
where $\gamma_e$ is the gyromagnetic ratio of the electron spin.

\begin{figure*}
\centering
\includegraphics[width=2\columnwidth]{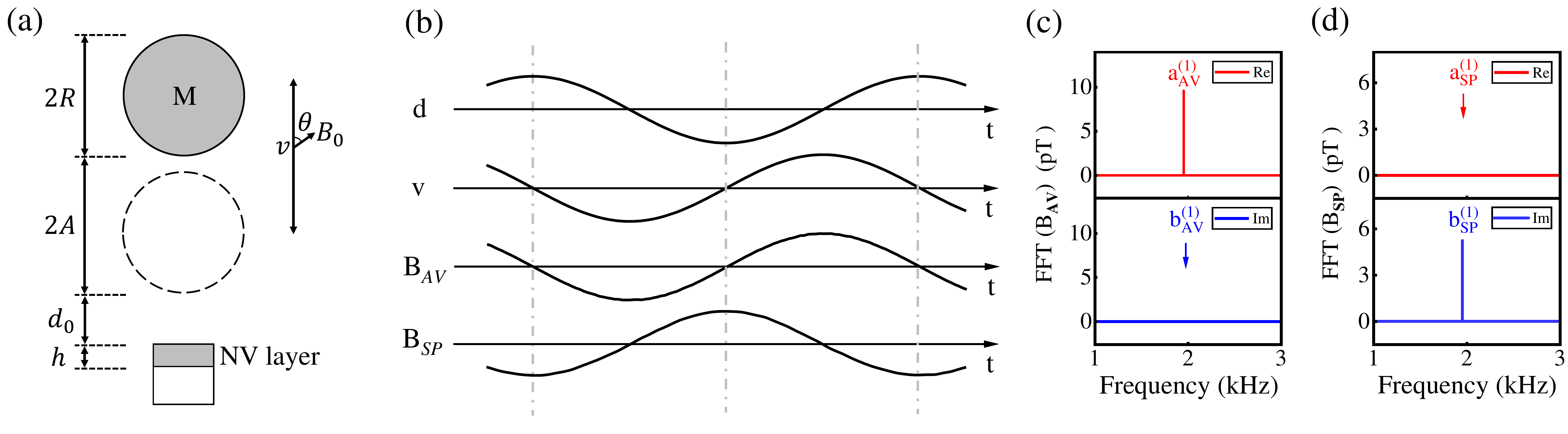}
\caption{Experimental detection scheme.
(a) Schematic experimental parameters. The lead sphere with the radius being R is denoted as M. The amplitude of the vibration is A,  and $d_0$  is the minimal distance between the bottom of M and the surface of the diamond. h is the thickness of layer with doped NV centers. v is the relative velocity vector between the sensor and the lead sphere. A static magnetic field $B_0$ is applied along the symmetry axis of NV centers. $\theta$ is the angle between $B_0$ and the velocity vector.
(b) The time evolution of the distance d(t), the velocity v(t), the simulated magnetic fields $B_{{AV}}(t)$ and $B_{SP}(t)$, respectively.  (c) and (d) are Fourier transform spectrum of $B_{{AV}}(t)$ and $B_{SP}(t)$, respectively.  For $B_{{AV}}$,  the none-zero component in the real part corresponds to $a_{{AV}}^{(1)}$ . For $B_{SP}$, the none-zero component in the imaginary part corresponds to  $b_{SP}^{(1)}$.
}
\label{figure1}
\end{figure*}

The detection scheme is shown in Fig. \ref{figure1}. The schematic diagram of nucleon source and NV ensemble are shown in Fig. \ref{figure1}(a).  The time-dependent distance between the bottom of the sphere and the surface of the diamond is $d(t) = d_0 + A [1+\cos (2\pi f_M t)]$, where $d_0$  is the minimal distance between the bottom of the sphere and the surface of the diamond. A and $f_M$ are the amplitude and frequency of the vibration, respectively. The velocity of the sphere can be written as $v (t) = - 2\pi f_MA\sin(2\pi f_M t)$. The possible effective magnetic field $B_{{AV}}$ and $B_{SP}$ sensed by the sensor can be estimated by integrating Eq. \eqref{BVA} and \eqref{BSP} over all nucleons in the sphere and electrons in the NV doped layer ( the detailed data and analysis have been included in the Appendix).
The time evolution of the effective magnetic field $B_{{AV}}(t)$ and $B_{SP}(t)$ is shown in Fig. \ref{figure1}(b), together with evolution of the distance d(t) and the velocity v(t). The effective magnetic field $B_{{AV}}$ and $B_{SP}$ can be decomposed into
\begin{equation}
B_{{AV}}= \sum_n [a_{{AV}}^{(n)} ~\sin(2\pi nf_Mt)  +  b_{{AV}}^{(n)} ~\cos(2\pi nf_Mt)],
\label{BeffAVt}
\end{equation}
\begin{equation}
B_{SP}= \sum_n [a_{SP}^{(n)} ~\sin(2\pi nf_Mt)  +  b_{SP}^{(n)} ~\cos(2\pi nf_Mt)],
\label{BeffSPt}
\end{equation}
where $a_{{AV}({SP})}^{(n)}$ and $b_{{AV}({SP})}^{(n)}$ are Fourier coefficients of the $n$-th harmonic.
We take $g_A^eg_V^N=1\times 10^{-20}$, $g_S^Ng_P^e = 10^{-20}$, $\lambda = 10^{-4}$ m  and $f_M = 1.953~$kHz as an example to numerically simulate the effective magnetic fields. The none-zero amplitudes of fourier series of $B_{{AV}}$ and $B_{SP}$ mainly lie in $a_{{AV}}^{(1)}$ and  $b_{SP}^{(1)}$ components, according to our experimental parameters \cite{SM}. The component with frequency being $1.953~$kHz  in the real (imaginary) part of Fourier transform spectrum corresponds to $a_{{AV}}^{(1)}$ ($b_{SP}^{(1)}$) as shown in Fig. \ref{figure1}(c) and (d). Thus our method can be utilized to explore the two types of exotic spin-dependent interactions simultaneously.

%The output of magnetometer was demodulated by the second lock-in amplifier (in Fig.\ref{figure1}a) with a reference signal $V_{ref} = V_{0} \cos (2\pi f_M t + \phi)]$, where $\phi = -32.4\pm8.8~^{\circ}$ was the experimental calibrated phase shift between output signal of magnetometer and d(t). With such detection method,  the in-phase and quadrature components of demodulated signal from the second lock-in amplifier are corresponding to $B_{{AV}}$ and  $B_{SP}$, respectively.

\begin{figure}
\centering
\includegraphics[width=1\columnwidth]{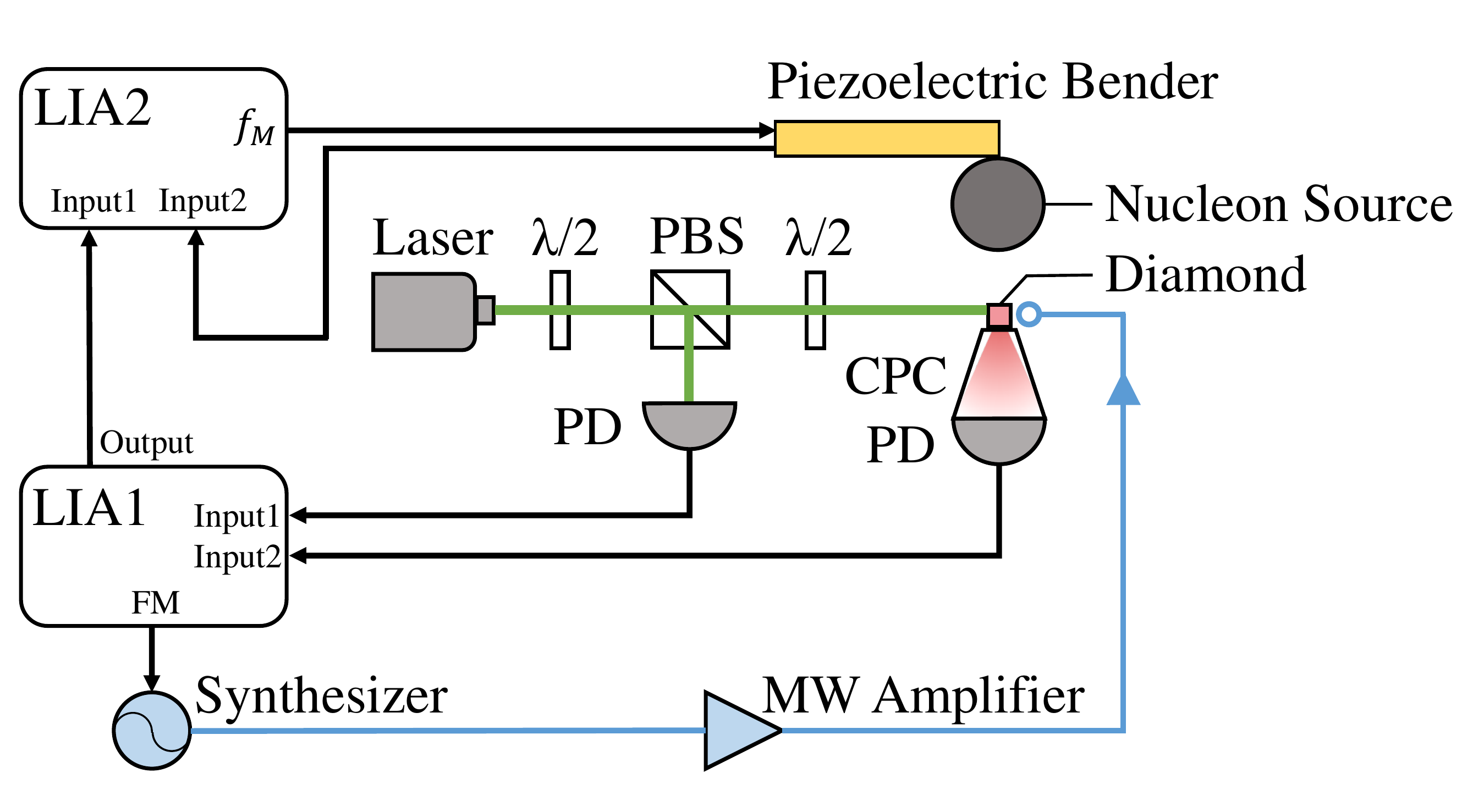}
\caption{ The schematic of our experimental setup. The 532-nm laser illuminates the layer of ensemble NV centers at the diamond surface. The red fluorescence is collected by a compound parabolic concentrator at the bottom.$\lambda$/2 denotes half-wave plate. PBS is the polarizing beam splitter. PD is the photodiode. CPC is the compound parabolic concentrator. LPF is the long-pass filter. FM is the modulation frequency of the microwave. $f_M$ is the modulation frequency of the nucleon source.
}
\label{figure2}
\end{figure}

Our experiment was conducted with an ensemble-NV-diamond magnetometer as shown in Fig. \ref{figure2}. A high-purity lead sphere with a radius of R = 978(3) $\mu$m was taken as the nucleon source with the density of nucleons being $6.8\times10^{30} ~m^{-3}$ \cite{Lee2018}. The lead sphere was attached to a piezoelectric bender, which can vibrate with a frequency $f_M=1.953 $~kHz. The magnetometer is an ensemble of NV centers in a 23-$\mu$m-thick layer at the surface of the diamond.
The substrate was a high purity electronic-grade $<$100$>$ oriented single crystal diamond with ppb nitrogen density. A 23-$\mu$m-thick nitrogen-rich layer was grown on the surface by chemical vapour deposition method. After electron irradiation and thermal annealing, a layer of NV ensemble with concentration of 14~ppm was obtained. A 532-nm laser with a diameter of 0.8~mm illuminated the NV doped layer via the flank of the diamond. The red fluorescence from the NV centers was collected by a compound parabolic concentrator \cite{Wolf2015}, filtered by a long-pass filter, and detected by a photodetector (PD). An external magnetic field of 20 Gauss along one of four symmetry axes of NV centers was applied.
Resonant microwave was applied to NV ensemble via a double-split-ring-resonator \cite{Bayat2014}. The fluorescence of NV ensemble varies as the external magnetic field changes, thus the NV ensemble can serve as a magnetometer. Recently, NV ensembles have been employed for high-sensitivity measurements of magnetic fields \cite{Zhang2021,Barry2017}.
In our experiment, to overcome low frequency noise, the frequency of microwave from the synthesizer was modulated with FM being 87.975 kHz. The signal of PD, which detected the fluorescence from NV centers, was demodulated by the first lock-in amplifier (LIA1) with demodulated frequency being $87.975$~kHz and the time constant being 8 $\mu$s. Another PD was used to monitor the power fluctuation of the laser for further noise cancellation. Then the output of the LIA1 was demodulated by the second lock-in amplifier (LIA2) with a reference signal $V_{ref} = V_{0} \cos (2\pi f_M t + \phi)]$, where $\phi = -32\pm 9~^{\circ}$ was the experimental calibrated phase shift between output signal of magnetometer and d(t). The specific information of the devices utilized in our setup and the calibration of the phase shift are included in Appendix. The time constant of LIA2 was set to be 10 ms. With such detection method,  the in-phase and quadrature components of demodulated signal from LIA2 are corresponding to $B_{{AV}}$ and  $B_{SP}$, respectively.

%Hereafter, the moving mass source and NV ensemble are denoted as M and S, respectively.

%We used Monte Carlo integration to numerically calculate $B_{{AV}}$ and $B_{SP}$ (see the Supplemental Material for details \cite{SM}).

\begin{figure}
\centering
\includegraphics[width=1\columnwidth]{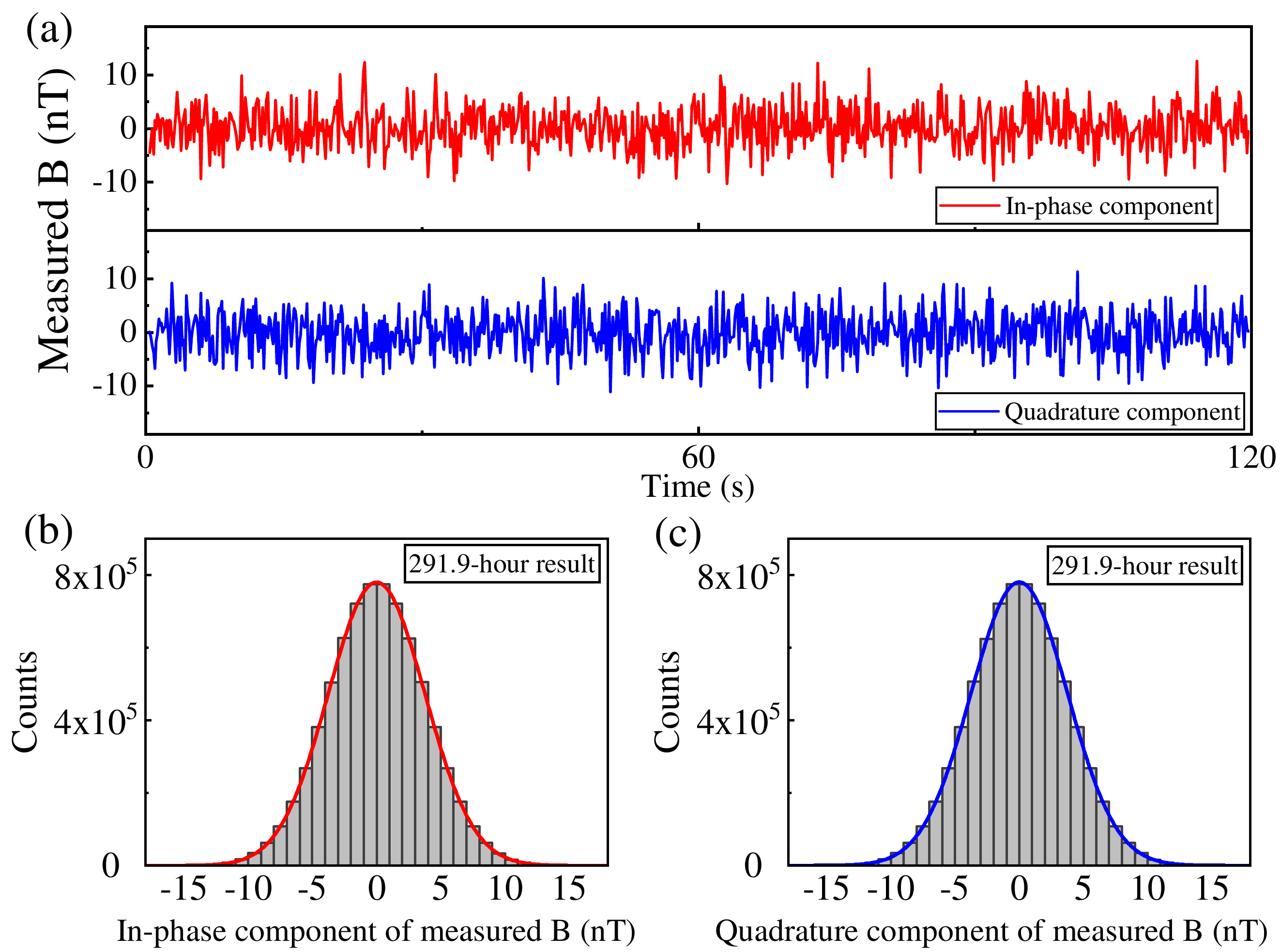}
\caption{ (a) Measurement of the two effective magnetic fields. The red (blue) line corresponds to $B_{{AV}}$ ($B_{SP}$) which is from the in-phase (quadrature) component of the output of the second lock-in amplifier. (b) and (c) The histograms of experimental results for 291.9-hour data. The red and blue solid lines are fits to the Gaussian distributions. The averages and the standard errors of the $B_{{AV}}$ and $B_{SP}$ are (0.1 $\pm$ 1.4)  pT  and  (-1.3 $\pm$ 1.4) pT, respectively.
}
\label{figure3}
\end{figure}

In our experiment, the sensitivity of the ensemble-NV-diamond magnetometer is 1.4~nT/Hz$^{1/2}$ within the frequency range from 0.4 to 2 kHz.
%The calibrated constant of magnetic field to output of magnetometer was 2.29(3)$\times 10^5$ V/T.
%A 1.953 kHz piezoelectric bender was chosen to drive the lead sphere.
The vibration amplitude was $A = 718(7)~$nm, and the minimal distance between the bottom of M and the surface of diamond was  $d_0 = 9.3(5)~\mu$m. In Fig. \ref{figure3}(a), the in-phase and quadrature parts of the output from LIA2 have been presented with the time duration being 120 seconds. The total experimental measurement was performed for 291.9 hours to suppress statistical fluctuations. The histograms of experimentally measured effective magnetic fields were shown in Fig. \ref{figure3}(b) and (c). The fits with Gaussian distribution to the histograms for 291.9-hour data provide the mean values and the standard errors of the effective magnetic fields.  The experimental $B_{{AV}}$ is (0.1 $\pm$ 1.4) pT, and the measured $B_{SP}$ is (-1.3 $\pm$ 1.4) pT. Our results show no evidence of the existence of exotic spin-dependent interactions, and experimental limits on both interactions can be obtained.

\begin{table}
	
	\caption{Summary of the systematic errors. The corrections to the constraint on $g_A^{e}g_V^{N}$ with $\lambda = 330~\mu$m  and $g_S^{N}g_P^{e}$ with $\lambda = 30~\mu$m are listed.}
	\label{table1}
      \linespread{5}
\renewcommand\arraystretch{1.2}
	\begin{tabular}{l c c c}

		\hline
		\hline
		  & & $\Delta g_A^{e}g_V^{N}$  & $\Delta g_S^{N}g_P^{e}$\\
             Parameter&Value&$(\times10^{-25})$  & $(\times10^{-21})$\\
		\hline
            Diamagnetism                  & $-1.6\times 10^{-5}$  & $\pm 0.3$  &  $\pm 2.9$ \\
		$ \theta $                  & $54.7\pm1.3^{\circ}$  & $^{+2.9}_{-2.8}$  &  $\pm 0.4$ \\
		Distance & $9.3\pm0.5~\mu$m      & $\pm 0.2$ & $\pm 0.4$\\
		Radius            & $978\pm 3~\mu$m     & $\pm 0.2$  & $\pm 0.3$\\
		Thickness  & $23\pm 1~\mu$m  &  $\pm 0.2$ &   $^{+0.3}_{-0.4}$ \\
		Amplitude   &$718\pm7~nm$&$^{+0.8}_{-1.0}$ &$^{+0.3}_{-0.4} $\\
		Deviation  &$0\pm10~\mu m$&$\pm 0.2$ &$^{+0.3}_{-0.4} $\\
		Phase delay $\phi$ &$-32\pm9^{\circ}$&$^{+2.6}_{-0.6}$ &$\pm 0.3 $\\
		Calib. Const. &$2.29\pm0.03\times10^5V/T$&$\pm 1.2 $ &$\pm 0.3 $\\
		\hline
\hline
		Final &&$\pm4.3$ &$\pm3.1$\\

		\hline
		\hline
	\end{tabular}
\end{table}

The systematic errors of our experiment are summarized in Table 1, where we take $\lambda = 330~\mu$m for $g_A^{e}g_V^{N}$ and $\lambda = 30~\mu$m for $g_S^{N}g_P^{e}$. The systematic errors in our experiment mainly come from uncertainties of parameters of the setup, such as the angle between the effective magnetic field and the NV axis, the distance between the bottom of M and the surface of the diamond, the radius of M and the vibration amplitude. We also consider the uncertainties of the thickness of NV layer and the misalignment between the center of diamond and the lead sphere in the x-y plane. Since our experiment has been performed in an external magnetic field being $ 20~$Gauss, one possible systematic error is due to the diamagnetism of M. The observable magnetic field variation due to diamagnetism of M is estimated to be less than 0.5 pT, which is smaller than standard error
so of the measured magnetic under the current statistics in our experiment. A detailed analysis of systematic errors is included in the Appendix.  Combing the systematic errors in quadrature, the total systematic error for $g_A^{e}g_V^{N}$ ($g_S^{N}g_P^{e}$) is derived to be $\pm 4.3\times10^{-25}$ ($\pm 3.1\times10^{-21}$). The bound for the coupling constant  $g_A^{e}g_V^{N}$ with $\lambda = 330~\mu$m is  $|g_A^{e}g_V^{N}|\leq 2.5\times 10^{-22}$ with  a 95$\%$ confidence level when both statistical and systematic errors are taken into account.  The limit for the coupling constant $g_S^{N}g_P^{e}$ with $\lambda = 30~\mu$m is  $|g_S^{N}g_P^{e}|\leq 2.5\times 10^{-20}$ with  a 95$\%$ confidence level. The other values of upper bound with different values of force range can de derived with the same procedure.

\begin{figure}
\centering
\includegraphics[width=1\columnwidth]{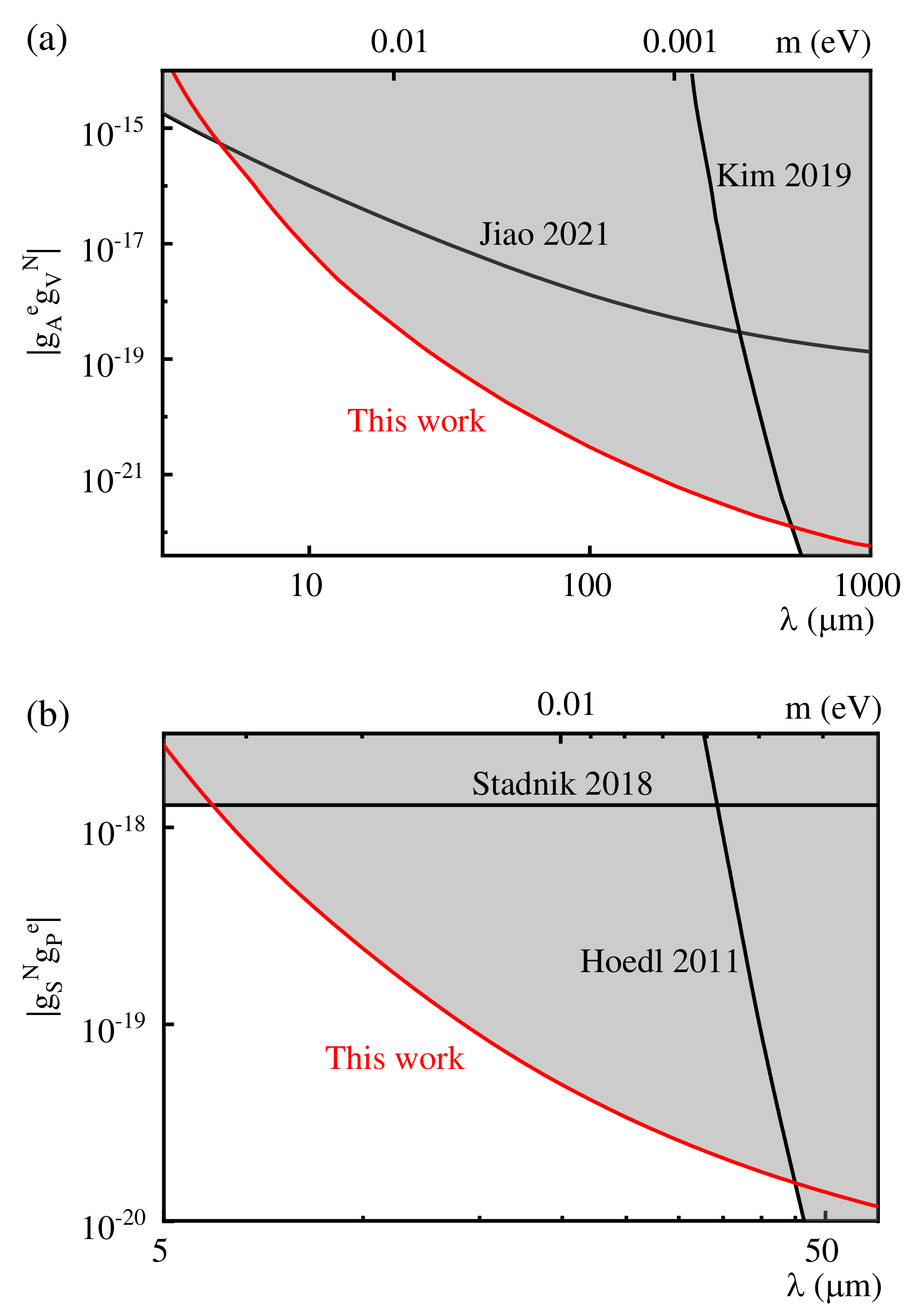}
\caption{(a) Upper limit on $g_A^eg_V^N $, as a function of the force range $\lambda$ and  mass of the bosons $m$. Black lines are upper limits established by experiments in Refs.\cite{Jiao2021PRL,Kim2019}. The red line is the upper bound obtained from our experiment, which establishes an improved laboratory bound in the force range from 5 to 500 $\mu$m. (b) Upper limit on $g_S^Ng_P^e$. Black lines are upper limits established by experiments in Refs.\cite{Stadnik2018,Hoedl2011}. Our experiment set the most stringent constraints in the force range from 5 to 45 $\mu$m.
}
\label{figure4}
\end{figure}

Fig. \ref{figure4}(a) shows the constraints on $g_A^eg_V^N $ set by this work together with limits established by previous experiments. The excluded values of coupling constant are presented as grey filled areas. The constraints of  $g_A^eg_V^N $ for the force range $\lambda < 5~\mu$m were established with single electron spin sensor by Jiao \emph{et~al.} \cite{Jiao2021PRL}. For force range $\lambda > 500~\mu$m, constraints were set by Kim \emph{et al}. \cite{Kim2019}, when an optically polarized vapor magnetometer was utilized to detect possible effective magnetic field from a BGO crystal. For the force range from 5 to 500$~\mu$m, the improved experimental limit is established by this work as the red line shown in Fig. \ref{figure4}(a). The upper bound for the force range $\lambda = 330~\mu$m,  is $|g_A^eg_V^N| \leq 2.5  \times10^{-22}$, which is more than three orders of magnitude more stringent than the bound established by the previous result \cite{Jiao2021PRL}.

As shown in Fig. \ref{figure4}(b), our work established improved constraints of $g_S^Ng_P^e$ in the force range from 6 to 45$~\mu$m. Recent experiments set limits with EDM experiments \cite{Stadnik2018} ($\lambda < 6~\mu$m) and  with torsion pendulum \cite{Hoedl2011} ($\lambda >45~\mu$m). The upper bound by our experiment at the force range $\lambda = 30~\mu$m is  $|g_S^Ng_P^e| \leq 2.5  \times10^{-20}$, which is more than one order of magnitude better than previous bound established by Ref.\cite{Stadnik2018}.

In summary, experimental searches for two types of exotic spin-dependent interactions have been performed by an ensemble-NV-diamond magnetometer. Improved constraints of two types of coupling constant have been established.  The current searching sensitivity is mainly limited by the sensitivity of the magnetometer. In future, the sensitivity of our magnetometer can be improved. Firstly, the concentration of NV centers in diamond can be optimized to enhance the coherence time \cite{Barry2020}. Secondly, the double resonance \cite{Fescenko2020} and the hyperfine structure \cite{Barry2017} could be used to further improve the sensitivity. Thirdly, other miscellaneous upgrades are helpful for the advancement, such as increasing the optical pumping rate \cite{Jensen2013}, using silicon carbide heatsink to cool the diamond to increase the signal contrast \cite{Schloss2018}. Our setup can also be utilized to search for other exotic spin-dependent interactions, such as an exotic parity-even spin- and velocity-dependent interaction between polarized electrons and nucleons, and interactions between polarized electrons. This work shows that an ensemble-NV-diamond magnetometer has great potential of exploring physics beyond the standard model.

This work was supported by the Chinese Academy of Sciences (Grants No. XDC07000000, No. GJJSTD20200001, No. QYZDY-SSW-SLH004, No.
QYZDB-SSW-SLH005), the National Key R$\&$D Program of China (Grant No. 2018YFA0306600, N0. 2021YFC2203100), Anhui Initiative in Quantum Information Technologies (Grant No. AHY050000) ,and NSFC (11961131007, 11653002).
X.\ R. thanks the Youth Innovation Promotion Association of Chinese Academy of Sciences for the support.
Y. \ F. \ C. and Y. \ W.  thank the Fundamental Research Funds for Central Universities.
Y. \ F. \ C. is supported in part by the CSC Innovation Talent Funds, by the USTC Fellowship for International Cooperation, and by the USTC Research Funds of the Double First-Class Initiative.

Hang Liang, Man Jiao and Yue Huang contributed equally to this work.

\appendix

\section {Numerical Simulations of the Effective magnetic fields}
	%ÃÉ¿¨¼ÆËã
	%¸µÀïÒ¶±ä»»
	In this section, we perform numerical simulations of the possible effective magnetic fields due to exotic spin-dependent interactions. The effective magnetic fields between polarized electron spin and nucleon are shown in Equ.(3) and (4) in the main text.
By Integrating over the volume of both the lead sphere and NV layer, we derive the possible effective magnetic fields $\textbf{B}_{ {AV}}$ and $\textbf{B}_{ {SP}}$ sensed by the NV ensemble as follows

	\begin{gather}
		\textbf{B}_{ {AV}}=\frac{1}{ {V_S}}\int_{ {V_S}}\int_{ {V_M}}\textbf{B}_{ {eff,AV}}(\textbf{r})\rho_M d {V_M} d {V_S},\\
		\textbf{B}_{ {SP}}=\frac{1}{ {V_S}}\int_{ {V_S}}\int_{ {V_M}}\textbf{B}_{ {eff,SP}}(\textbf{r})\rho_M d {V_M} d {V_S},
	\end{gather}
where $\rho_M=6.8\times10^{30} ~m^{-3}$ is the nucleon density of the lead sphere. $V_S$ and $V_M$ are integral volume of NV layer and lead sphere, respectively.
The radius of the lead sphere is $R=978(3)~\mu m$ . The size of NV layer is $ 660\times661\times23~\mu m^3$. The minimal distance $d_0$ is 9.3(5) $\mu$m.
	\par The Monte Carlo method can be utilized to numerically calculate the effective magnetic fields to avoid complex calculations due to high integral dimensions \cite{Kim2019}. The algorithm of Monte Carlo integral is performed as follows:
	\begin{itemize}
		\item[(1)] $N_{ {MC}}=2^{20}$ random pairs of points inside both the volumes of the lead sphere and the NV ensemble are generated.
		\item[(2)] The effective magnetic field $B_{ {eff,AV}}^i$ ($B_{ {eff,SP}}^i$) between a randomly generated pair of points is calculated with a given force range.
		\begin{gather}
			B_{ {eff,AV}}^i=\frac{g^e_Ag^N_V}{2\pi\gamma_e}\frac{e^{-\frac{r}{\lambda}}}{r} v cos\theta,\\
              B_{ {eff,SP}}^i=g^N_Sg^e_P\frac{\hbar}{4\pi m_e\gamma_e}(\frac{1}{\lambda r}+\frac{1}{r^2})e^{-\frac{r}{\lambda}} \frac{z}{r} cos\theta,
		\end{gather}
		where $\theta= arccos(1/\sqrt{3})$ is the angle between the direction of the velocity $\textbf{v}$ and the NV axis.
z is the component of distance vector $\textbf{r}$ along the direction of the velocity $\textbf{v}$ .
		\item[(3)] All the contributions to the effective magnetic field are summed and normalized to give the average magnetic field generated by the lead sphere and sensed by the NV ensemble:\\
		\begin{gather}
			B_{ {AV}}=N_{ {nucleon}}\frac{1}{N_{ {MC}}} \sum_{i}^{N_{ {MC}}} B_{ {eff,AV}}^i,\\
              B_{ {SP}}=N_{ {nucleon}}\frac{1}{N_{ {MC}}} \sum_{i}^{N_{ {MC}}} B_{ {eff,SP}}^i,\\
		\end{gather}
		where $N_{ {nucleon}}$ is the total number of nucleons in lead sphere.
	\end{itemize}
	
The magnetic field $B_{ {AV}}$ and $B_{ {SP}}$ can be decomposed into orthogonal components of Fourier series as shown in Equ.(5) and (6) in the main text, the coefficients can be derived as
         \begin{gather}
			a_{ {AV(SP)}}^{(n)}=\frac{2}{T}\int_0^T\sin(2\pi nf_Mt)B_{ {AV(SP)}}(t)dt,\\
              b_{ {AV(SP)}}^{(n)}=\frac{2}{T}\int_0^T\cos(2\pi nf_Mt)B_{ {AV(SP)}}(t)dt,
		\end{gather}

We take ${g^e_Ag^N_V}= 1\times {10}^{-20}$, and $\lambda={10}^{-4}~m$ as an example, the coefficients of $B_{ {AV}}$ at first three harmonic frequencies are listed as follows

\begin{center}
\renewcommand\arraystretch{1.5}
\begin{tabular}{|c|c|c|c|}

\hline
n&1&2&3 \\
\hline
$a_{ {AV}}^{(n)}$ (pT)~&9.62&0.02&0.00\\
\hline
$b_{ {AV}}^{(n)}$ (pT)~&0&0&0\\
\hline

\end{tabular}

\end{center}
The amplitude of the first harmonic coefficient  $a_{ {AV}}^{(1)}$ is much larger than higher order harmonic coefficients $a_{ {AV}}^{(2)}$ and $a_{ {AV}}^{(3)}$.  Values of $b_{ {AV}}^{(n)}$ are zero. Similarly, the calculated coefficients of $B_{ {SP}}$ at the first three harmonic frequencies are

\begin{center}
\renewcommand\arraystretch{1.5}
\begin{tabular}{|c|c|c|c|}
\hline
n&1&2&3 \\
\hline
$a_{ {SP}}^{(n)}$(pT)& 0 & 0&0\\
\hline
$b_{ {SP}}^{(n)}$(pT)& 5.24 & -0.06&-0.06\\
\hline

\end{tabular}
\end{center}
when ${g^N_Sg^e_P}= 1\times {10}^{-20}$, and $\lambda={10}^{-4}~m$. The amplitude of the first harmonic coefficient $b_{ {SP}}^{(1)}$ is much larger than higher order harmonic coefficients $b_{ {SP}}^{(2)}$ and $b_{ {SP}}^{(3)}$, and values of $a_{ {SP}}^{(n)}$ are zero.
The possible effective magnetic fields mainly lie in the first order harmonic coefficients with the experimental parameters in our setup.
%The output signal of ensemble-NV-diamond magnetometer was demodulated with LIA2 with a reference signal $V_{ {ref}}=V_0 \cos(2\pi f_Mt +\phi)$ and the time constant of LIA2 is 10 ms to extract the first coefficients of $B_{ {AV}}$ and $B_{ {SP}}$ (ie, $a_{ {AV}}^{(1)}$ and $b_{ {SP}}^{(1)}$) .

%
%In our experiment, the output signal of ensemble-NV-diamond magnetometer was demodulated with a lock-in amplifier(LIA2) with a reference signal $V_{ref}=V_0 cos(2\pi f_Mt +\phi)$. The phase delay $\phi$ between the output signal of the magnetometer and feedback of piezoelectric bender was calibrated to guarantee demodulation at the same phase of $d(t)$. Therefore the in-phase component of the LIA demodulated signal was used to detect $B_{SP}$ and only the signal at the first harmonic frequency was preserved. The quadrature component of the LIA demodulated signal was used to detect $B_{AV}$ and only the signal at the first harmonic frequency was preserved. We can use Y output to detect $B_{AV}$ and X output to detect $B_{SP}$ at the same time.

%	\noindent\textbf{simulation of the effective magnetic fields generated by the Piezoelectric Bimorph}

      \section{Calibration of the phase delay}
The phase delay $\phi$  between the output signal of the ensemble-NV-diamond magnetometer and $d(t)$ can be calibrated by a given signal as discussed in Ref. \cite{Su2021}. A thin copper wire carrying a DC current was stuck to the front section of the piezoelectric bender. The magnetic field generated by the copper wire was modulated by the vibration of the piezoelectric bender. In our experiment, a calibration magnetic field oscillating with an amplitude of (18 $\pm$ 2) nT was applied, and the phase delay $\phi$ was obtained as $\phi = (-32 \pm 9)~ ^{\circ}$.

%	\begin{figure}[http]
%		\centering
%		\includegraphics[width=1\columnwidth]{FigS1.pdf}
%		\caption{\textbf{Calibration of the phase delay.}(a) Experimental setup used  to calibrate phase delay. (b) Schematic of phase delay between $d(t)$ and output of magnetometer.}
%		\label{FigS3}
%	\end{figure}
%

	\section {The performance of the ensemble-NV-diamond magnetometer}

	The ensemble-NV-diamond magnetometer used in our experiment was based on the continuous-wave (CW) method, wherein laser and microwave fields are continuously applied to NV centers. The NV centers of $|m_s=0\rangle$ state can be transmitted to $|m_s=+1\rangle$ by a resonance microwave with an angular frequency $\omega_e=D + \gamma_e B_0 $, where $D=2\pi\times2.87$~GHz is the ground-state zero splitting, $\gamma_e=2\pi\times28$~GHz/T is the gyromagnetic ratio of the electron spin, and $B_0$ is the external magnetic field along the symmetry axis of NV centers generated by a solenoid coil. When the external magnetic field varies, the population on $|m_s=+1\rangle$ states decreases, resulting in changes in fluorescence which could be detected.

In order to avoid flicker noise, the frequency modulation technique was used in our experiment. The frequency of microwave from the synthesizer was modulated by a lock-in amplifier(LIA1 in Fig. 2 of the main text) with FM being 87.975~kHz, and the signal of PD, which detected the fluorescence from NV centers was demodulated with the same frequency. For further noise cancellation, the signal of a reference PD used to monitor the power fluctuation of the laser was also demodulated by LIA1 with a frequency of 87.975~kHz.
The laser intensity noise was canceled by scaling and subtracting the demodulated reference signal from the demodulated fluorescence signal, with a cancellation coefficient of about 2 \cite{Schloss2018}.

\begin{figure}[http]
		\centering
		\includegraphics[width=1\columnwidth]{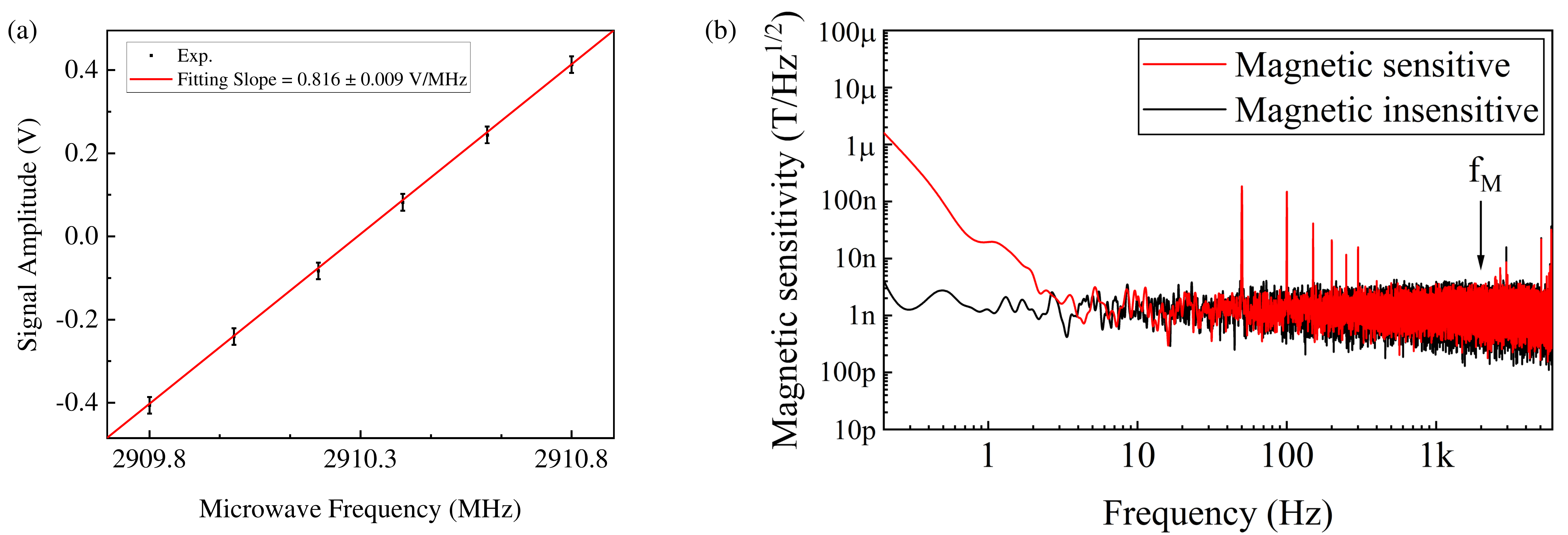}
		\caption{\textbf{The performance of the magnetometer.}(a) The CW spectrum. The red line is the linear fitting to give the max slope. (b) Magnetic sensitivity of ensemble-NV-diamond magnetometer. The peaks of red line at 50~Hz and harmonics are the magnetic noise due to power supply of the equipment. The vibration frequency of lead sphere $f_M$ = 1.953 kHz is also displayed.}
		\label{FigS1}
	\end{figure}

The calibrated constant of magnetic field to output of magnetometer was determined by the max slope of the continuous-wave spectrum, which was 0.816 $\pm$ 0.009 V/MHz as shown in Fig. \ref{FigS1}(a), and corresponds to a calibrated constant of $2.29 \pm 0.03 \times 10^5 $~V/T with the gyromagnetic ratio $\gamma_e=2\pi\times28$~GHz/T. The sensitivity of 1.4 nT/Hz$^{1/2}$ from 0.4 to 2 kHz was achieved, as shown in Fig. \ref{FigS1}(b).

%	\section {Experimental results and error analysis}
%	\noindent\textbf{Experimental results}
%
%	\begin{figure}[http]
%		\centering
%		\includegraphics[width=0.5\columnwidth]{Fig.S5.pdf}
%		\caption{\textbf{the histogram of experimental data}}
%		\label{FigS5}
%	\end{figure}
%
%
%	We accumulated 290 hours of experimental data with sampling rate of 7$Hz$ to extract the constraints on the coupling strength $g^N_Vg^e_A$ and $g^N_Sg^e_P$. A gaussian fitting to the histogram of experimental data gives the mean measured magnetic field and standard error of $B_{AV}$ is (0.1 $\pm$ 1.4)$ \times  10^{-12}$ T, and the result of the measured magnet field $B_{SP}$ is (-1.3 $\pm$ 1.4) $\times  10^{-12}$ T.
%	
%	The coupling strength can be given by
%	\begin{gather}
%		g^N_Vg^e_A=\frac{B_{AV}}{a^{1}_{AV}} \\
%		g^N_Sg^e_P=\frac{B_{SP}}{b^{1}_{SP}}
%	\end{gather}
%	where $a^{1}_{AV}$ is the first cosine harmonic coefficient corresponding to ${g^e_Ag^N_V}^{sim}= 1$ (see Sec.III), $b^{1}_{SP}$ is the first sinusoidal harmonic coefficient corresponding to ${g^N_Sg^e_P}^{sim}= 1$ (see Sec.III). Considering the uncertainty propagation, we can obtain
%	\begin{gather}
%		g^e_Ag^N_V (\lambda=330\mu\text{m})=(0.09\pm{1.32}_{stat})\times {10}^{-22} \\
%		g^N_Sg^e_P (\lambda=30\mu\text{m})=(0.79\pm{0.85}_{stat})\times {10}^{-20}
%	\end{gather}
%	
%	
%	

	   \section {List of experiment Instruments}
\begin{table*}[]
\caption{Detailed information about the devices in our experimental setup. }
\label{tab1}
\begin{tabular}{lll}
\hline
\textbf{Instrument}&\textbf{Company}& \textbf{Model}\\
\hline
Lock-in Amplifier1 and 2 & Zurich Instruments & HF2LI\\
\hline
Laser&Cobolt&  0532-05-01-1500-700  \\
\hline
Synthesizer&National Instruments&FSW-0010\\
\hline
MW Amplifier&CIQTEK&GYPA2530-42\\
\hline
PD&Thorlabs&SM05PD1A\\
\hline
Piezoelectric Bender&Harbin Core Tomorrow&NAC2223\\
\hline
\end{tabular}
\end{table*}

The schematic of experimental setup has been shown in Fig. 2 in the main text. Table \ref{tab1} shows the companies and models of devices used in our experiment.

	\section{Systematic Error Analysis}
	\noindent\textbf{Diamagnetism of the lead sphere}

	In the external magnetic field of $B_0 ~=~20$~Gauss,  the diamagnetism of the lead sphere causes a magnetic field $B_ {diam}$ on NV ensemble.
    \begin{equation}
\mathbf{B}_{ {diam}}=\frac{1}{V_ {S}}\int_{V_ {S}}  ~\int_{V_ {M}} \frac{\chi}{4 \pi}\left[\frac{3 \mathbf{r}\left(\mathbf{B}_{0} \cdot \mathbf{r}\right)}{r^{5}}-\frac{\mathbf{B}_{0}}{r^{3}}\right] d V_ {M} d V_ {S},
\label{B_diam}
\end{equation}
where $\chi=-16\times10^{-6}$,  is  magnetic susceptibility of the lead sphere \cite{CRC}. $V_ {M}$ is the integral volume of lead sphere, $V_ {S}$ is the volume of NV ensemble.

\begin{figure}[http]
		\centering
		\includegraphics[width=0.75\columnwidth]{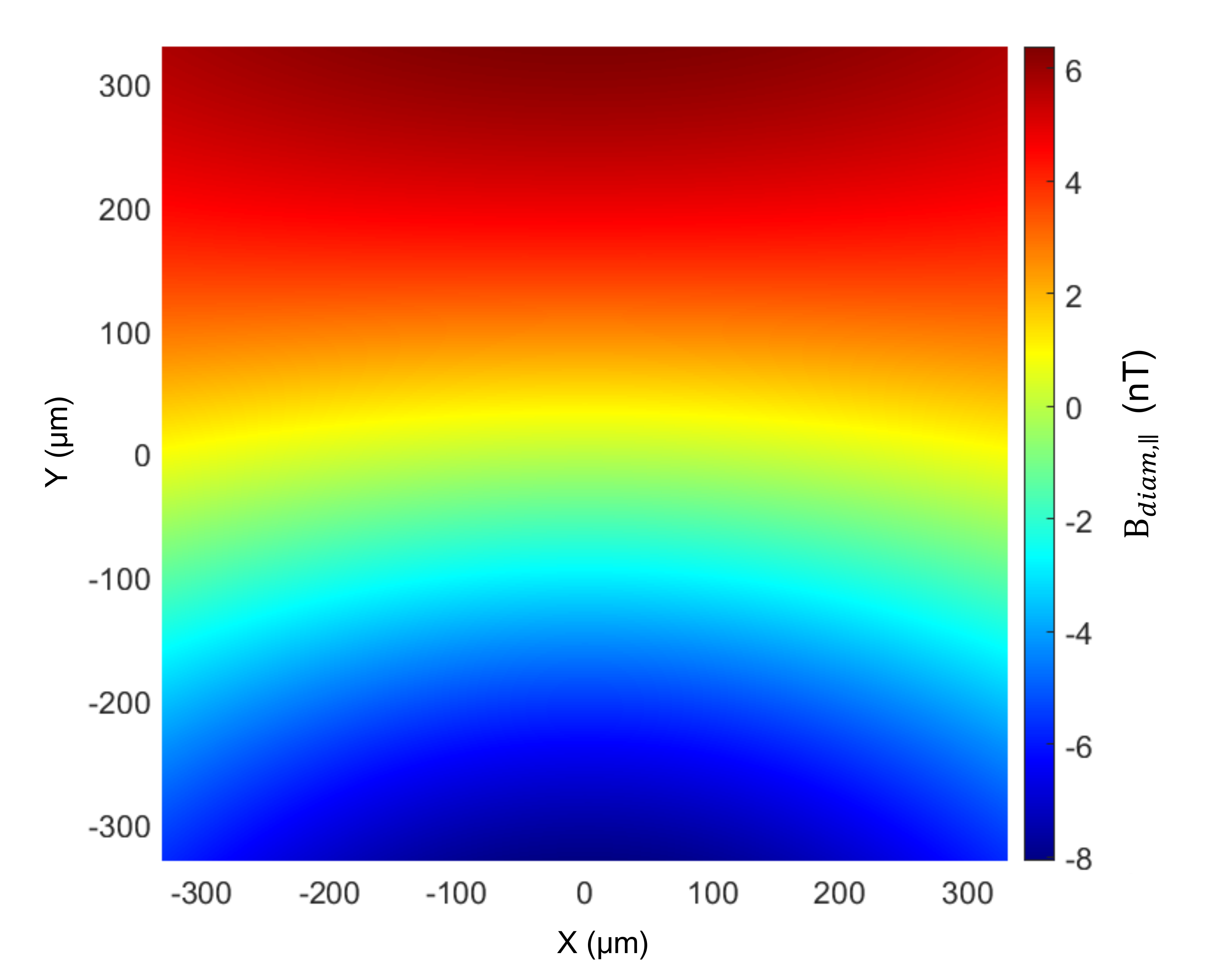}
		\caption{\textbf{Contour plot of $B_{diam,\parallel}$  on NV ensemble in x-y plane.}}
		\label{FigBdiam}
	\end{figure}

Since the large zero filed splitting of NV centers, the magnetic field perpendicular to NV axis can be ignored.
The magnetic field parallel to NV axis due to diamagnetism is denoted as $B_ {diam,\parallel}$. The distribution of $B_ {diam,\parallel}$ in NV ensemble is shown in Fig. \ref{FigBdiam}.
The x-y plane is perpendicular to the vibration of piezoelectric bender, x-axis stands for the direction perpendicular to the NV axis.

As shown in Fig. \ref{FigBdiam}, $B_ {diam,\parallel}$ has a sign reversal along y-axis.
Thus the averaged magnetic field sensed by the NV ensemble due to diamagnetism is in the range from 0.738~pT to 0.740~pT during the vibration of lead sphere,  corresponding to a magnetic field of 0.002~pT, which is much less than the standard error of the measured field under current statistics. The possible misalignment of the lead sphere and diamond is also taken into consideration. The misalignment is estimated to be $\pm10 ~\mu m$, and the maximum variation of magnetic field due to diamagnetism is 0.5~pT, which is taken as a systematic error. The correction to the constraint on $g^e_Ag^N_V$ is (0.0$\pm0.3)\times 10^{-25}$ at $\lambda=330~ \mu$m. The correction to the constraint on $g^N_Sg^e_P$ is (0.0$\pm2.9)\times 10^{-21}$ at $\lambda=30~ \mu$m.

	\noindent\textbf{Uncertainty in $d_0$}

	The minimal distance between the bottom of the lead sphere and the diamond $d_0$ is measured to be 9.3(5)~$\mu$m. To estimate the corrections to the constraints on $g^e_Ag^N_V$  and $g^N_Sg^e_P$ due to the uncertainty in $d_0$ , ${10}^{5}$ samples for $d_0$ was randomly taken, which satisfied a Gaussian distribution $P_{d_0}(d_{0,i})=\frac{1}{\sqrt{2 \pi} \sigma_{d_0}} exp[-\frac{(d_{0,i}-\mu_{d_0})^2}{2 {\sigma_{d_0}}^2}]$.  $\mu_{d_0}=9.3~\mu$m and $\sigma_{d_0}=0.5~\mu$m are the mean value and uncertainty of measured $d_0$.  The correction to $g^e_Ag^N_V $ is ~$(0.0 \pm 0.2)\times 10^{-25}$ at $\lambda=330~ \mu$m. The correction to $g^N_Sg^e_P$ is $(0\pm0.4)\times 10^{-21}$ at $\lambda=30~ \mu$m.
	
	\noindent\textbf{Uncertainty in $R$}

	The radius of the lead sphere is measured to be $R=978(3)~\mu$m. The corrections to the constraints on $g^e_Ag^N_V$ and $g^N_Sg^e_P$ are in the same procedure as for the uncertainty in $d_0$. The correction to the constraint on $g^e_Ag^N_V$ is (0.0$\pm0.2)\times 10^{-25}$ at $\lambda=330~ \mu$m. The correction to  the constraint on$g^N_Sg^e_P$ is (0.0$\pm0.3)\times 10^{-21}$ at $\lambda=30~ \mu$m.
	
		\noindent\textbf{Uncertainty in $\theta$}

	The uncertainty in $\theta$ is estimated to be 1.3$^{\circ}$. The correction to the constraint on $g^e_Ag^N_V$ is from $ -2.8 \times 10^{-25}$ to $2.9 \times 10^{-25}$ at $\lambda=330~ \mu$m. The correction to the constraint on $g^N_Sg^e_P$ is (0.0$\pm0.4)\times 10^{-21}$ at $\lambda=30~ \mu$m.

	\noindent\textbf{Uncertainty in $h$}

	The thickness of the NV layer is estimated by measuring the thickness of the diamond before and after the growth of NV layer. The original thickness of the diamond is 551(1)~$\mu m$. After growth of the NV layer, the thickness of the diamond is 574(1)~$\mu m$. The thickness of the NV layer is measured to be $h=23(1)~\mu$m. The correction to the constraint on $g^e_Ag^N_V$ is (0.0$\pm0.2)\times 10^{-25}$ at $\lambda=330~ \mu$m. The correction to  the constraint on $g^N_Sg^e_P$ is  from $-0.4 \times 10^{-21}$  to $0.3 \times 10^{-21}$ at $\lambda=30 ~\mu$m.

	\noindent\textbf{Uncertainty in $A$}

	The vibration amplitude is measured to be $A=718(7)$~nm. The correction to the constraint on $g^e_Ag^N_V$ is from $-1.0\times 10^{-25}$  to $0.8\times10^{-25}$  at $\lambda=330~ \mu$m. The correction to the constraint on $g^N_Sg^e_P$ is $-0.4\times 10^{-21}$  to $0.3\times10^{-21}$  at $\lambda=30 ~\mu$m.

	\noindent\textbf{Deviation in x-y plane.}

	The deviation in x-y plane is measured to be $(0\pm 10)~\mu$m.  The correction to  the constraint on  $g^e_Ag^N_V$ is (0.0$\pm0.2)\times 10^{-25}$ at $\lambda=330~ \mu$m. The correction to the constraint on $g^N_Sg^e_P$ is $-0.4\times 10^{-21}$  to $0.3\times10^{-21}$  at $\lambda=30 ~\mu$m.

	\noindent\textbf{Uncertainty in $\phi$}

	The phase delay is measured to be $\phi=(-32 \pm 9 )$~degree. The correction to  the constraint on $g^e_Ag^N_V$ is from $-0.6 \times 10^{-25}$ to $ 2.6\times 10^{-25}$ at $\lambda=330~ \mu$m. The correction to  the constraint on  $g^N_Sg^e_P$ is (0.0$\pm0.3)\times 10^{-21}$ at $\lambda=30~ \mu$m.

	\noindent\textbf{Uncertainty in Calib. Const.}

	The Calib. Const.  is measured to be $(2.29\pm 0.03 )\times {10}^{5}$ V/T. The correction to  the constraint on $g^e_Ag^N_V$ is (0.0$\pm1.2)\times 10^{-25}$ at $\lambda=330~ \mu$m. The correction to  the constraint on  $g^N_Sg^e_P$ is (0.0$\pm0.3)\times 10^{-21}$ at $\lambda=30~ \mu$m.

	With the fiducial probability of  95$\%$, the upper bound $|g_A^eg_V^N| \leq 2.5  \times10^{-22}$ for the force range $\lambda$~=~330$~\mu$m and the upper bound $|g_S^Ng_P^e \leq 2.5  \times10^{-20}$ for the force range $\lambda$ = 30~$\mu$m were obtained, taking both statistical and systematic errors into account.

\end{document}